# Efficiently determining Convergence in Polynomial Recurrence Sequences

Deepak Ponvel Chermakani

deepakc@pmail.ntu.edu.sg   deepakc@e.ntu.edu.sg   deepakc@ed-alumni.net   deepakc@myfastmail.com   deepakc@usa.com

*Abstract: -*   We derive the necessary and sufficient condition, for a given Polynomial Recurrence Sequence to converge to a given target rational K. By converge, we mean that the Nth term of the sequence, is equal to K, as N tends to positive infinity. The basic idea of our approach is to construct a univariate polynomial equation in x, whose coefficients correspond to the terms of the Sequence. The approach then obtains the condition by analyzing five cases that cover all possible real values of x. The condition can be evaluated within time that is a polynomial function of the size of the description of the Polynomial Recurrence Sequence, hence convergence or non-convergence can be efficiently determined.

## 1. Introduction

There has been a lot of study [1][2][3] into the convergence properties of linear recurrence sequences and polynomial recurrence sequences. Some authors have focussed on whether the value of the $N^{th}$ term of the sequence (not necessary an integer sequence), can asymptotically converge to some real, as $N$ tends to infinity. Other authors have focussed on whether the ratio of the $N^{th}$ term to the $(N+1)^{th}$ term can asymptotically converge to some real, as $N$ tends to infinity.

   In this paper, we develop an approach to determine whether or not the $N^{th}$ term of a given sequence, can become equal to a given target rational $K$, as $N$ tends to infinity. The starting points and the coefficients of the sequence, are rationals. The term "rational" denotes a real number $(x/y)$ where both $x$ and $y$ are integers and $y \neq 0$. In the rest of this paper, when we say $p_i$ "converges to $K$", we mean that the value of $p_N = K$, as $N$ tends to infinity. In this paper, by "infinity" we mean "positive infinity". We also denote the absolute value function of $x$ as $abs(x)$, so $abs(x) = x$ if $x \geq 0$, and $abs(x) = -x$ if $x < 0$.

   Sections 2 and 3 apply our approach to homogeneous linear recurrence sequences and non-homogeneous linear recurrence sequences, respectively. Section 4 applies our approach to polynomial recurrence sequences

## 2. Homogeneous Linear Recurrence Sequences

Consider a homogeneous linear recurrence sequence as follows:
   $p_i = c_i$ , for all integers $i$ in $[0, L-1]$ , and
   $p_i = p_{i-1} a_1 + p_{i-2} a_2 + ... + p_{i-L+1} a_{L-1} + p_{i-L} a_L$ , for all integers $i \geq L$, where,
$< c_0 , c_1 , c_2 , ... , c_{L-1} >$ and $< a_1 , a_2 , ... , a_L >$ are given vectors of rationals, where $a_L \neq 0$, and where $p_i$ denotes the $i^{th}$ term in our sequence.

   Define the univariate Polynomial $A(x) = ( 1 - a_1 x - a_2 x^2 - ... - a_L x^L )$, and the Polynomial $P(x) = ( p_0 + p_1 x + p_2 x^2 + ... + p_N x^N )$. Denoting $A(x) P(x)$ as $R(x)$, we have $R(x) =$
$p_0 + (p_1 - p_0 a_1)x + (p_2 - p_1 a_1 - p_0 a_2)x^2 + (p_3 - p_2 a_1 - p_1 a_2 - p_0 a_3)x^3 + ... + (p_{L-1} - p_{L-2} a_1 - p_{L-3} a_2 - ... - p_0 a_{L-1})x^{L-1}$
$+ (p_L - p_{L-1} a_1 - p_{L-2} a_2 - ... - p_0 a_L)x^L + ... + (p_N - p_{N-1} a_1 - p_{N-2} a_2 - ... - p_{N-L} a_L)x^N$
$+ ( - p_N a_1 - p_{N-1} a_2 - ... - p_{N-L+1} a_L)x^{N+1}$
$+ ( - p_N a_2 - ... - p_{N-L+2} a_L)x^{N+2}$
$+ ( - p_N a_3 - ... - p_{N-L+3} a_L)x^{N+3}$
...
$+ ( - p_N a_L)x^{N+L}$

   It may be noted that the value of $( p_i - p_{i-1} a_1 - p_{i-2} a_2 - ... - p_0 a_i )x^i$ is zero for all integers $i$ in $[L,N]$, since each $p_i$ is a term of our sequence. Hence $R(x) =$
$p_0 + (p_1 - p_0 a_1)x + (p_2 - p_1 a_1 - p_0 a_2)x^2 + (p_3 - p_2 a_1 - p_1 a_2 - p_0 a_3)x^3 + ... + (p_{L-1} - p_{L-2} a_1 - p_{L-3} a_2 - ... - p_0 a_{L-1})x^{L-1}$
$+ ( - p_N a_1 - p_{N-1} a_2 - ... - p_{N-L+1} a_L)x^{N+1}$

$+ (- p_N a_2 - ... - p_{N-L+2} a_L) x^{N+2}$

$+ (- p_N a_3 - ... - p_{N-L+3} a_L) x^{N+3}$

...

$+ (- p_N a_L) x^{N+L}$

## 2.1 Convergence to zero

**Theorem-1:** ( $p_i$ converges to $0$ ) $\leftrightarrow$ ( $c_i = 0$ for all integers $i$ in $[0, L-1]$ )

**Proof:** We show two methods of proving this.

**Proof-Method-1:** Rewrite the recurrence in terms of $p_{i-L}$ as follows:

$p_{i-L} = p_i / a_L - p_{i-1} (a_1 / a_L) - p_{i-2} (a_2 / a_L) - p_{i-3} (a_3 / a_L) - ... - p_{i-L+1} (a_{i-L+1} / a_L)$

So assuming that $p_N = p_{N-1} = ... = p_{N-L+1} = 0$ as $N$ tends to infinity, then $p_{N-L} = 0$. This means that ( $p_i$ converges to $0$ ) $\rightarrow$ ( $p_N = p_{N-1} = ... = p_2 = p_1 = p_0 = 0$) $\rightarrow$ ( $c_i = 0$ for all integers $i$ in $[0, L-1]$ ). It is obvious that ( $c_i = 0$ for all integers $i$ in $[0, L-1]$ ) $\rightarrow$ ( $p_L = p_{L+1} = ... = p_N = 0$) $\rightarrow$ ( $p_i$ converges to $0$ ). Thus ( $p_i$ converges to $0$ ) $\leftrightarrow$ ( $c_i = 0$ for all integers $i$ in $[0, L-1]$ ).

**Proof-Method-2:** If A and B are two boolean statements, then (A$\leftrightarrow$B) can be proved by proving (B$\rightarrow$A) and (A$\rightarrow$B). If $p_i = 0$ for all integers $i$ in $[0, L-1]$, it is trivial to see that $p_i = 0$ for all integers $i \geq L$. Hence (B$\rightarrow$A). Assume that $p_i$ converges to $0$. Then for sufficiently large $N$, $R(x) = p_0 + (p_1 - p_0 a_1)x + (p_2 - p_1 a_1 - p_0 a_2)x^2 + (p_3 - p_2 a_1 - p_1 a_2 - p_0 a_3)x^3 + ... + (p_{L-1} - p_{L-2} a_1 - p_{L-3} a_2 - ... - p_0 a_{L-1})x^{L-1}$. Since the degree of $A(x)$ is $L$, $P(x)$ is a normal Polynomial (i.e. not a rational-Polynomial), and the degree of $R(x)$ is $(L-1)$, the only way for $A(x)P(x) = R(x)$ is that $p_i = 0$, for all integers $i$ in $[0,N]$, which happens when $p_i = 0$ for all integers $i$ in $[0, L-1]$. This implies $c_i = 0$ for all integers $i$ in $[0, L-1]$. Hence (A$\rightarrow$B).

**Hence Proved**

## 2.2 Convergence to a non-zero target rational

In this section, we derive the sufficient and necessary conditions for $p_i$ to converge to a non-zero target rational $K$.

**Theorem-2:** ( $p_i$ converges to $K$ where $K \neq 0$ ) $\leftrightarrow$ ( ( $c_i = K$ for all integers $i$ in $[0, L-1]$ ) AND ( $(a_1 + a_2 + ... + a_{L-1} + a_L) = 1$) )

**Proof:** We show two methods of proving this.

**Proof-Method-1:** Rewrite the recurrence in terms of $p_{i-L}$ as follows:

$p_{i-L} = p_i / a_L - p_{i-1} (a_1 / a_L) - p_{i-2} (a_2 / a_L) - p_{i-3} (a_3 / a_L) - ... - p_{i-L+1} (a_{i-L+1} / a_L)$

Assuming that $p_N = p_{N-1} = ... = p_{N-L+1} = K$ as $N$ tends to infinity where $K$ is non-zero, then ($p_{N-L} = K$) if $(1 = 1/a_L - a_1/a_L - a_2/a_L - a_3/a_L - ... - a_{i-L+1}/a_L)$, i.e. if $(a_1 + a_2 + ... + a_L = 1)$. This means that ($p_i$ converges to $K$ ) $\rightarrow$ (($p_N = p_{N-1} = ... = p_1 = p_0 = K$) AND $(a_1 + a_2 + ... + a_L = 1)$) $\rightarrow$ (($c_i = K$ for all integers $i$ in $[0, L-1]$) AND $(a_1 + a_2 + ... + a_L = 1)$). Next, it is also obvious that (($c_i = K$ for all integers $i$ in $[0, L-1]$) AND $(a_1 + a_2 + ... + a_L = 1)$) $\rightarrow$ ($p_0 = p_1 = ... = p_{N-1} = p_N = K$) $\rightarrow$ ($p_i$ converges to $K$). Thus ($p_i$ converges to $K$ where $K \neq 0$) $\leftrightarrow$ (($c_i = K$ for all integers $i$ in $[0, L-1]$) AND $(a_1 + a_2 + ... + a_{L-1} + a_L = 1)$).

**Proof-Method-2:** If A and B are two boolean statements, then (A$\leftrightarrow$B) can be proved by proving (B$\rightarrow$A) and (A$\rightarrow$B). If ( $c_i = K$ for all integers $i$ in $[0, L-1]$ ) AND ( $(a_1 + a_2 + ... + a_{L-1} + a_L) = 1$), it is trivial to see that $p_i = K$ for all integers $i \geq L$. Hence (B$\rightarrow$A). We shall now proceed to prove that (A$\rightarrow$B). Assume that $N$ is extremely large and $p_N = p_{N-1} = ... = p_{N-L} = K$. So $A(x)P(x) = R(x) = p_0 + (p_1 - p_0 a_1)x + (p_2 - p_1 a_1 - p_0 a_2)x^2 + (p_3 - p_2 a_1 - p_1 a_2 - p_0 a_3)x^3 + ... + (p_{L-1} - p_{L-2} a_1 - p_{L-3} a_2 - ... - p_0 a_{L-1})x^{L-1}$

$+ K(- a_1 - a_2 - ... - a_L)x^{N+1}$

$+ K(- a_2 - ... - a_L)x^{N+2}$

$+ K(- a_3 - ... - a_L)x^{N+3}$

...

$+ K(- a_L)x^{N+L}$

Our aim is to prove that there exists a real Polynomial $P(x)$, such that for every real $x$, $A(x)P(x) = R(x)$. For this, we consider five cases ($x=0$, $0 < abs(x) < 1$, $abs(x) > 1$, $x=1$, and $x=-1$). The union of the conditions obtained in all five cases, will be necessary and sufficient, for $p_i$ to converge to $K$. We proceed to analyse each case in detail.

<u>CASE-1 ($x = 0$)</u>: $A(0)=1$, $P(0)=p_0$, and $R(0)=p_0$. Hence, no new condition is extracted from CASE-1.

<u>CASE-2 ($0 \leq abs(x) < 1$)</u>: Let $M$ be the minimum integer, such that $p_i = K$, for all $i \geq M$. Note that $N$ is the degree of $P(x)$ and is a much larger number. So as $N$ tends to infinity, the value of $x^N$ tends to $0$ because $x < 1$, as a result of which $R(x) = p_0 + (p_1 -$

$p_0 a_1)x + (p_2 - p_1 a_1 - p_0 a_2)x^2 + (p_3 - p_2 a_1 - p_1 a_2 - p_0 a_3)x^3 + \ldots + (p_{L-1} - p_{L-2} a_1 - p_{L-3} a_2 - \ldots - p_0 a_{L-1})x^{L-1}$. Also, $P(x) = p_0 + p_1 x + p_2 x^2 + \ldots + p_M x^M + K x^{M+1} + K x^{M+2} + \ldots + K x^N$. We also know that the sum $1 + x^2 + x^3 + \ldots + x^N$ as $N$ tends to infinity, is equal to $1/(1-x)$, if $x<1$. So $P(x) = p_0 + p_1 x + p_2 x^2 + \ldots + p_M x^M + (K x^{M+1}/(1-x)) = (p_0 + (p_1 - p_0)x + (p_2 - p_1) x^2 + \ldots + (p_M - p_{M-1}) x^M + (K - p_M) x^{M+1}) / (1 - x)$. So we have $(1-x)A(x)P(x) = A(x)(p_0 + (p_1 - p_0)x + (p_2 - p_1) x^2 + \ldots + (p_M - p_{M-1}) x^M + (K - p_M) x^{M+1})$, and we have $(1-x)R(x) = (1-x) (p_0 + (p_1 - p_0 a_1)x + (p_2 - p_1 a_1 - p_0 a_2)x^2 + (p_3 - p_2 a_1 - p_1 a_2 - p_0 a_3)x^3 + \ldots + (p_{L-1} - p_{L-2} a_1 - p_{L-3} a_2 - \ldots - p_0 a_{L-1})x^{L-1})$. Since the degree of $(1-x)R(x)$ is at most $L$, and the degree of $A(x)$ is $L$, the degree of $(p_0 + (p_1 - p_0)x + (p_2 - p_1) x^2 + \ldots + (p_M - p_{M-1}) x^M + (K - p_M) x^{M+1})$ has to be 0. This implies $p_M = p_{M-1} = \ldots = p_0 = K$. Plugging these values, we have $(1-x)A(x)P(x) = (1 - a_1 x - a_2 x^2 - \ldots - a_L x^L)$. We also have $(1-x)R(x) = K(1-x)(1 + (1 - a_1)x + (1 - a_1 - a_2)x^2 + (1 - a_1 - a_2 - a_3)x^3 + \ldots + (1 - a_1 - a_2 - \ldots - a_{L-1})x^{L-1}) = K(1 - a_1 x - a_2 x^2 - \ldots - a_{L-1} x^{L-1} - (1 - a_1 - a_2 - \ldots - a_{L-1}) x^L)$. Since the coefficients of $x^L$ in both $(1-x)A(x)P(x)$ and $(1-x)R(x)$ need to be equal, we need $(1 - a_1 - a_2 - \ldots - a_{L-1}) = a_L$, which implies $(a_1 + a_2 + \ldots + a_{L-1} + a_L = 1)$. So the condition extracted from CASE-2 is $((c_i = K$ for all integers $i$ in $[0,L-1]$ ) AND $(a_1 + a_2 + \ldots + a_{L-1} + a_L = 1))$.

CASE-3 (*abs(x) > 1*): In $A(x)P(x) = R(x)$, divide throughout by $x^{N+L}$, and let $y = 1/x$, so that we have:
$(y^L - a_1 y^{L-1} - a_2 y^{L-2} - \ldots - a_{L-2} y^2 - a_{L-1} y - a_L) (p_0 y^N + p_1 y^{N-1} + p_2 y^{N-2} + \ldots + p_{N-2} y^2 + p_{N-1} y + p_N) =$
$p_0 y^{N+L} + (p_1 - p_0 a_1)y^{N+L-1} + (p_2 - p_1 a_1 - p_0 a_2)y^{N+L-2} + (p_3 - p_2 a_1 - p_1 a_2 - p_0 a_3)y^{N+L-3} + \ldots + (p_{L-1} - p_{L-2} a_1 - p_{L-3} a_2 - \ldots - p_0 a_{L-1})y^{N+1}$
$+ (-p_N a_1 - p_{N-1} a_2 - \ldots - p_{N-L+1} a_L) y^{L-1}$
$+ (-p_N a_2 - \ldots - p_{N-L+2} a_L) y^{L-2}$
$+ (-p_N a_3 - \ldots - p_{N-L+3} a_L) y^{L-3}$
...
$+ (-p_N a_{L-1} - p_{N-1} a_L) y$
$+ (-p_N a_L)$

Utilizing the condition from CASE-2 that $p_i = K$, for all integers $i \geq 0$, and since $N$ tends to infinity, $A(x)P(x) = K(y^L - a_1 y^{L-1} - a_2 y^{L-2} - \ldots - a_{L-2} y^2 - a_{L-1} y - a_L ) / (1-y)$, and $R(x) = K((-a_1 - a_2 - \ldots - a_L)y^{L-1} + (-a_2 - \ldots - a_L)y^{L-2} + (-a_3 - \ldots - a_L)y^{L-3} + \ldots + (-a_{L-1} - a_L)y + (-a_L))$. So $(1-y)A(x)P(x) = K(y^L - a_1 y^{L-1} - a_2 y^{L-2} - \ldots - a_{L-2} y^2 - a_{L-1} y - a_L)$, and $(1-y)R(x) = K((a_1 + a_2 + \ldots + a_L)y^L - a_1 y^{L-1} - a_2 y^{L-2} - \ldots - a_{L-2} y^2 - a_{L-1} y - a_L) = K(y^L - a_1 y^{L-1} - a_2 y^{L-2} - \ldots - a_{L-2} y^2 - a_{L-1} y - a_L)$. So $A(x)P(x) = R(x)$. Hence, no new condition is extracted from CASE-3.

CASE-4 *(x=1)*: We have $A(1)P(1) = (1 - a_1 - a_2 - \ldots - a_L) (p_0 + p_1 + p_2 + \ldots + p_{N-2} + p_{N-1} + p_N)$. Utilizing the condition from CASE-2, we have $A(1)P(1) = 0$. Also, $R(1) =$
$p_0 + (p_1 - p_0 a_1) + (p_2 - p_1 a_1 - p_0 a_2) + (p_3 - p_2 a_1 - p_1 a_2 - p_0 a_3) + \ldots + (p_{L-1} - p_{L-2} a_1 - p_{L-3} a_2 - \ldots - p_0 a_{L-1})$
$+ K(-a_1 - a_2 - \ldots - a_L)$
$+ K(-a_2 - \ldots - a_L)$
$+ K(-a_3 - \ldots - a_L)$
...
$+ K(-a_L)$.

Utilizing the condition from CASE-2, we have $R(1) =$
$K( \quad 1 + (1 - a_1) + (1 - a_1 - a_2) + (1 - a_1 - a_2 - a_3) + \ldots + (1 - a_1 - a_2 - \ldots - a_{L-1})$
$\quad + (-a_1 - a_2 - \ldots - a_L)$
$\quad + (-a_2 - \ldots - a_L)$
$\quad + (-a_3 - \ldots - a_L)$
$\quad \ldots$
$\quad + (-a_L) \quad )$
$= KL(1 - a_1 - a_2 - \ldots - a_L) = 0$.

Since $A(1)P(1) = R(1)$, there is no new condition extracted from CASE-4.

CASE-5 *(x=-1)*: We have $A(-1)P(-1) = (1 + a_1 - a_2 + a_3 \ldots -(-1)^L a_L) (p_0 - p_1 + p_2 - p_3 \ldots + (-1)^N p_N)$, and $R(-1) = p_0 - (p_1 - p_0 a_1) + (p_2 - p_1 a_1 - p_0 a_2) - (p_3 - p_2 a_1 - p_1 a_2 - p_0 a_3) + \ldots + (-1)^{L-1} (p_{L-1} - p_{L-2} a_1 - p_{L-3} a_2 - \ldots - p_0 a_{L-1})$
$+ (-1)^N K ((a_1 + a_2 + \ldots + a_L)$
$- (a_2 + a_3 \ldots + a_L)$
$+ (a_3 + a_4 + \ldots + a_L)$
$- (a_4 + a_5 \ldots + a_L)$

...
$+ (-1)^L (a_L))$

Utilizing the condition from CASE-2 that $p_i = K$, for all integers $i \geq 0$, we have $R(-1) = K (1 - (1 - a_1) + (1 - a_1 - a_2) - (1 - a_1 - a_2 - a_3) + ... + (-1)^{L-1} (1 - a_1 - a_2 - ... - a_{L-1})$
$+ (-1)^N K ((a_1 + a_2 + ... + a_L)$
$- (a_2 + a_3 ... + a_L)$
$+ (a_3 + a_4 + ... + a_L)$
$- (a_4 + a_5 ... + a_L)$
...
$+ (-1)^L (a_L)))$

We now consider 4 sub-cases, for $N$ and $L$ each being odd or even.

Subcase-5.1 (L is even, N is even): $A(-1)P(-1) = K(1 + a_1 - a_2 + a_3 ... - a_L)$, while $R(-1) = K(a_1 + a_3 + a_5 + ... + a_{L-1}$
$+ a_1 + a_3 + a_5 + ... + a_{L-1})$. Utilizing the condition from CASE-4 that $(1 - a_1 - a_2 - ... - a_L) = 0$, we have $R(-1) = K(1 + a_1 - a_2 + a_3 ... - a_L)$. Hence, no new condition can be extracted from this Subcase.

Subcase-5.2 (L is even, N is odd): $P(-1) = 0$, so $A(-1)P(-1) = 0$. $R(-1) = K(a_1 + a_3 + a_5 + ... + a_{L-1}$
$- a_1 - a_3 - a_5 - ... - a_{L-1}) = 0$. Hence, no new condition can be extracted from this Subcase.

Subcase-5.3 (L is odd, N is even): $A(-1) P(-1) = K(1 + a_1 - a_2 + a_3 ... + a_L)$, while $R(-1) = K(1 - a_1 - a_2 - ... - a_{L-1}$
$+ (a_1 + a_3 + a_5 + ... + a_L)) = K(1 + a_1 - a_2 + a_3 ... + a_L)$. Hence, no new condition can be extracted from this Subcase.

Subcase-5.4 (L is odd, N is odd): $P(-1) = 0$, so $A(-1)P(-1) = 0$. $R(-1) = K(1 - a_2 - a_4 - ... - a_{L-1} - (a_1 + a_3 + ... a_L))$, which as per the condition from CASE-4, is $0$. Hence, no new condition can be extracted from this Subcase.

Hence, no new condition can be extracted from CASE-5.

The condition extracted from all five CASES, is $((p_i = c_i = K,$ for all integers $i$ in $[0, L-1])$ AND $((a_1 + a_2 + ... + a_{L-1} + a_L) = 1))$. Hence (A→B).

**Hence Proved**

## 3. Non-Homogeneous Linear Recurrence Sequences

Consider a non-homogeneous linear recurrence sequence as follows:

$q_i = c_i$, for all integers $i$ in $[0,L-1]$, and

$q_i = q_{i-1} a_1 + q_{i-2} a_2 + ... + q_{i-L+1} a_{L-1} + q_{i-L} a_L + d$, for all integers $i \geq L$, where,

$d$ is a non-zero rational, $<c_0, c_1, c_2, ..., c_{L-1}>$ and $<a_1, a_2, ..., a_L>$ are given vectors of rationals, where $a_L \neq 0$, and where $q_i$ denotes the $i^{th}$ term in our sequence.

Define the univariate Polynomial $A(x) = (1 - a_1 x - a_2 x^2 - ... - a_L x^L)$, the Polynomial $B(x) = (dx^L + dx^{L+1} + ... + dx^{N-1} + dx^N)$, and the Polynomial $P(x) = (q_0 + q_1 x + q_2 x^2 + ... + q_N x^N)$, where $N$ tends to infinity. Denoting $R(x) = (A(x)P(x) - B(x))$, we have $R(x) = q_0 + (q_1 - q_0 a_1)x + (q_2 - q_1 a_1 - q_0 a_2)x^2 + (q_3 - q_2 a_1 - q_1 a_2 - q_0 a_3)x^3 + ... + (q_{L-1} - q_{L-2} a_1 - q_{L-3} a_2 - ... - q_0 a_{L-1})x^{L-1}$
$+ (q_L - q_{L-1} a_1 - q_{L-2} a_2 - ... - q_0 a_L - d)x^L + ... + (q_N - q_{N-1} a_1 - q_{N-2} a_2 - ... - q_{N-L} a_L - d)x^N$
$+ (- q_N a_1 - q_{N-1} a_2 - ... - q_{N-L+1} a_L)x^{N+1}$
$+ (- q_N a_2 - ... - q_{N-L+2} a_L)x^{N+2}$
$+ (- q_N a_3 - ... - q_{N-L+3} a_L)x^{N+3}$
...
$+ (- q_N a_L)x^{N+L}$

It may be noted that the value of $(q_i - q_{i-1} a_1 - q_{i-2} a_2 - ... - q_{i-L} a_i - d)x^i$ is zero for all integers $i$ in $[L,N]$, since each $q_i$ is a term of our sequence. Also, as $N$ is large, we have $q_N = q_{N-1} = ... = q_{N-L} = K$. Hence $R(x) =$
$q_0 + (q_1 - q_0 a_1)x + (q_2 - q_1 a_1 - q_0 a_2)x^2 + (q_3 - q_2 a_1 - q_1 a_2 - q_0 a_3)x^3 + ... + (q_{L-1} - q_{L-2} a_1 - q_{L-3} a_2 - ... - q_0 a_{L-1})x^{L-1}$
$+ K(- a_1 - a_2 - ... - a_L)x^{N+1}$

$+ K(-a_2 - ... - a_L)x^{N+2}$

$+ K(-a_3 - ... - a_L)x^{N+3}$

...

$+ K(-a_L)x^{N+L}$

### 3.1 Convergence to zero

**Theorem-3:** $q_i$ **cannot converge to** $0$

**Proof:** We show two methods of proving this.

**Proof-Method-1:** Assume $q_i$ converges to $0$. Then as $N$ tends to infinity, $q_{N-1} = q_{N-2} = ... = q_{N-L} = 0$, but then $q_N = d$, which contradicts our assumption. Hence, $q_i$ cannot converge to $0$.

**Proof-Method-2:** Assume $q_i$ converges to $0$. As $N$ tends to infinity, $R(x) =$
$q_0 + (q_1 - q_0 a_1)x + (q_2 - q_1 a_1 - q_0 a_2)x^2 + (q_3 - q_2 a_1 - q_1 a_2 - q_0 a_3)x^3 + ... + (q_{L-1} - q_{L-2} a_1 - q_{L-3} a_2 - ... - q_0 a_{L-1})x^{L-1}$
This means $B(x)+R(x) = q_0 + (q_1 - q_0 a_1)x + (q_2 - q_1 a_1 - q_0 a_2)x^2 + (q_3 - q_2 a_1 - q_1 a_2 - q_0 a_3)x^3 + ... + (q_{L-1} - q_{L-2} a_1 - q_{L-3} a_2 - ... - q_0 a_{L-1})x^{L-1} + dx^L + ... + dx^N$. Also, $A(x)P(x) = B(x)+R(x)$. However, $q_N = q_{N-1} = q_{N-2} = ... = q_{N-L} = 0$, which means $A(x)P(x) = 0$. And $B(x)+R(x) \neq 0$ because $d \neq 0$. This contradicts our assumption. Hence, $q_i$ cannot converge to $0$.

**Hence Proved**

**Theorem-4:** ( $q_i$ **converges to** $K$ **where** $K \neq 0$ ) $\leftrightarrow$ ( ( $c_i = K$ **for all integers** $i$ **in** $[0, L-1]$ ) **AND** ( $(a_1 + a_2 + ... + a_{L-1} + a_L) = (1-(d/K))$ ) )

**Proof:** We show two methods of proving this.

**Proof-Method-1:** Rewrite the recurrence in terms of $q_{i-L}$ as follows:

$q_{i-L} = q_i / a_L - q_{i-1}(a_1/a_L) - q_{i-2}(a_2/a_L) - q_{i-3}(a_3/a_L) - ... - q_{i-L+1}(a_{i-L+1}/a_L) - d/a_L$

Assuming that $q_N = q_{N-1} = ... = q_{N-L+1} = K$ as $N$ tends to infinity where $K$ is non-zero, then $(q_{N-L} = K)$ if $(1 = 1/a_L - a_1/a_L - a_2/a_L - a_3/a_L - ... - a_{i-L+1}/a_L - d/(Ka_L))$, i.e. if $(a_1 + a_2 + ... + a_L = (1-(d/K)))$. This means that $(q_i$ converges to $K) \to ((q_N = q_{N-1} = ... = q_1 = q_0 = K)$ AND $(a_1 + a_2 + ... + a_L = (1-(d/K)))) \to ((c_i = K$ for all integers $i$ in $[0, L-1])$ AND $(a_1 + a_2 + ... + a_L = (1-(d/K))))$. Next, it is also obvious that $((c_i = K$ for all integers $i$ in $[0, L-1])$ AND $(a_1 + a_2 + ... + a_L = (1-(d/K)))) \to (q_0 = q_1 = ... = q_{N-1} = q_N = K) \to (q_i$ converges to $K)$. Thus $(q_i$ converges to $K$ where $K \neq 0) \leftrightarrow ((c_i = K$ for all integers $i$ in $[0, L-1])$ AND $(a_1 + a_2 + ... + a_{L-1} + a_L = (1-(d/K))))$.

**Proof-Method-2:** If A and B are two boolean statements, then (A$\leftrightarrow$B) can be proved by proving (B$\to$A) and (A$\to$B). If $((c_i = K$ for all integers $i$ in $[0, L-1]$ ) AND ( $(a_1 + a_2 + ... + a_{L-1} + a_L) = (1-(d/K))))$, it is trivial to see that $q_i = K$ for all integers $i \geq L$. Hence (B$\to$A). We shall now proceed to prove that (A$\to$B). Assume $q_i$ converges to $K$. As $N$ tends to infinity, $R(x) =$
$q_0 + (q_1 - q_0 a_1)x + (q_2 - q_1 a_1 - q_0 a_2)x^2 + (q_3 - q_2 a_1 - q_1 a_2 - q_0 a_3)x^3 + ... + (q_{L-1} - q_{L-2} a_1 - q_{L-3} a_2 - ... - q_0 a_{L-1})x^{L-1}$

$+ K(-a_1 - a_2 - ... - a_L)x^{N+1}$

$+ K(-a_2 - ... - a_L)x^{N+2}$

$+ K(-a_3 - ... - a_L)x^{N+3}$

...

$+ K(-a_L)x^{N+L}$

We also have $(A(x)P(x) - B(x)) = ( 1 - a_1 x - a_2 x^2 - ... - a_L x^L)(q_0 + q_1 x + q_2 x^2 + ... + q_{M-1} x^{M-1} + K(x^M + x^{M+1} + ... + x^{N-1} + x^N)) - (dx^L + dx^{L+1} + ... + dx^{N-1} + dx^N)$. Here $M$ is the minimum integer for which $q_i = K$, for all $i \geq M$.

Our aim is to prove that there exists a real Polynomial $P(x)$, such that for every real $x$, $(A(x)P(x) - B(x)) = R(x)$. For this, we consider five cases ($x=0$, $0 < abs(x) < 1$, $abs(x) > 1$, $x=1$, and $x=-1$). The union of the conditions obtained in all five cases, will be necessary and sufficient, for $q_i$ to converge to $K$. We proceed to analyse each case in detail.

<u>CASE-1</u> ($x = 0$): $A(0)=1$, $P(0)=q_0$, $B(0)=0$ and $R(0)=q_0$. So $(A(0)P(0) - B(0)) = R(0) = q_0$. Hence, no new condition is extracted from CASE-1.

<u>CASE-2</u> ($0 \leq abs(x) \leq 1$): In this case, $(A(x)P(x) - B(x)) = ( 1 - a_1 x - a_2 x^2 - ... - a_L x^L)(q_0 + q_1 x + q_2 x^2 + ... + q_{M-1} x^{M-1} + Kx^M/(1-x)) - dx^L/(1-x)$. So $(1-x)(A(x)P(x) - B(x)) = ( 1 - a_1 x - a_2 x^2 - ... - a_L x^L)(q_0 + (q_1 - q_0)x + (q_2 - q_1) x^2 + ... (q_{M-1} - q_{M-2})x^{M-1} + (K - q_{M-1})x^M) - dx^L$. And $(1-x)R(x) = (1-x)( q_0 + (q_1 - q_0 a_1)x + (q_2 - q_1 a_1 - q_0 a_2)x^2 + (q_3 - q_2 a_1 - q_1 a_2 - q_0 a_3)x^3 + ... + (q_{L-1} - q_{L-2} a_1 - q_{L-3} a_2 - ... - q_0 a_{L-1})x^{L-1}$ ). Since, the degree of $(1-x)R(x)$ is at most $L$, the degree of $(1-x)P(x)$ has to be $0$. This means $q_{M-1} = q_{M-2} = ... = q_1 = q_0 = K$. Plugging these values, we have $(1-x)(A(x)P(x) - B(x)) = K( 1 - a_1 x - a_2 x^2 - ... - a_L x^L ) - dx^L$. We also have $(1-x)R(x) = K(1-x)( 1 + (1 - a_1)x + (1 - a_1 - a_2)x^2 + (1 - a_1 - a_2 - a_3)x^3 + ... + (1 - a_1 - a_2 - ... - a_{L-1})x^{L-1} ) = K( 1 - a_1 x - a_2 x^2 - ... - a_{L-1} x^{L-1} - (1 - a_1 - a_2 - ... - a_{L-1})x^L )$. So comparing the coefficients of $x^L$ in both $(1-x)(A(x)P(x) - B(x))$ and $(1-x)R(x)$, we need to have $K( 1 - a_1 - a_2 - ... - a_{L-1} ) = Ka_L + d$. So the condition extracted from CASE-2 is ( ( $c_i = K$ for all integers $i$ in $[0,L-1]$ ) AND ($a_1 + a_2 + ... + a_{L-1} + a_L = 1-(d/K)$) ).

CASE-3 (abs(x) > 1): In $(A(x)P(x)-B(x)) = R(x)$, divide throughout by $x^{N+L}$ and let $y = 1/x$, so that we have:
$(y^L - a_1 y^{L-1} - a_2 y^{L-2} - ... - a_{L-2} y^2 - a_{L-1} y - a_L )(q_0 y^N + q_1 y^{N-1} + q_2 y^{N-2} + ...+ q_{N-2} y^2 + q_{N-1} y + q_N) - d(y^L + y^{L+1} +...+ y^{N-1} + y^N) =$
$q_0 y^{N+L} + (q_1 - q_0 a_1)y^{N+L-1} + (q_2 - q_1 a_1 - q_0 a_2)y^{N+L-2} + (q_3 - q_2 a_1 - q_1 a_2 - q_0 a_3)y^{N+L-3} + ... + (q_{L-1} - q_{L-2} a_1 - q_{L-3} a_2 - ... - q_0 a_{L-1})y^{N+1}$
$+ ( - q_N a_1 - q_{N-1} a_2 - ... - q_{N-L+1} a_L) y^{L-1}$
$+ ( - q_N a_2 - ... - q_{N-L+2} a_L) y^{L-2}$
$+ ( - q_N a_3 - ... - q_{N-L+3} a_L) y^{L-3}$
...
$+ ( - q_N a_{L-1} - q_{N-1} a_L) y$
$+ ( - q_N a_L)$

Utilizing the conditions from CASE-2, and since $N$ tends to infinity, the $(A(x)P(x)-B(x)) = (K( y^L - a_1 y^{L-1} - a_2 y^{L-2} - ... - a_{L-2} y^2 - a_{L-1} y - a_L ) - dy^L ) / ( 1-y )$, and the $R(x)= K( ( - a_1 - a_2 - ... - a_L)y^{L-1} + ( - a_2 - ... - a_L)y^{L-2} + ( - a_3 - ... - a_L)y^{L-3} + ... + ( - a_{L-1} - a_L)y + ( - a_L) )$. So $(1-y)(A(x)P(x)-B(x)) = (K( y^L - a_1 y^{L-1} - a_2 y^{L-2} - ... - a_{L-2} y^2 - a_{L-1} y - a_L ) - dy^L ) = K( (1-(d/K))y^L - a_1 y^{L-1} - a_2 y^{L-2} - ... - a_{L-2} y^2 - a_{L-1} y - a_L )$, and $( 1-y )R(x) = K(( a_1 + a_2 + ... + a_L)y^L - a_1 y^{L-1} - a_2 y^{L-2} - ... - a_{L-2} y^2 - a_{L-1} y - a_L ) = K( (1-(d/K))y^L - a_1 y^{L-1} - a_2 y^{L-2} - ... - a_{L-2} y^2 - a_{L-1} y - a_L )$. So $(A(x)P(x)-B(x))=R(x)$. Hence, no new condition is extracted from CASE-3.

CASE-4 (x=1): We have $(A(1)P(1)-B(1)) = (1 - a_1 - a_2 - ... - a_L) ( q_0 + q_1 + q_2 + ...+ q_{N-2} + q_{N-1} + q_N) - d(N-L+1)$. Utilizing the condition from CASE-2, we have $(A(1)P(1)-B(1)) = (K(N+1)(1 - a_1 - a_2 - ... - a_L) - d(N-L+1)) = (N+1)(K(1 - a_1 - a_2 - ... - a_L ) - d) + dL = K(N+1)(1 - (d/K) - a_1 - a_2 - ... - a_L ) + dL = dL$. Also, $R(1) = q_0 + (q_1 - q_0 a_1) + (q_2 - q_1 a_1 - q_0 a_2) + (q_3 - q_2 a_1 - q_1 a_2 - q_0 a_3) + ... + (q_{L-1} - q_{L-2} a_1 - q_{L-3} a_2 - ... - q_0 a_{L-1})$
$+ K( - a_1 - a_2 - ... - a_L)$
$+ K( - a_2 - ... - a_L)$
$+ K( - a_3 - ... - a_L)$
...
$+ K( - a_L)$.

Utilizing the condition from CASE-2, we have $R(1) =$
$K($
$\quad 1 + (1 - a_1) + (1 - a_1 - a_2) + (1 - a_1 - a_2 - a_3) + ... + (1 - a_1 - a_2 - ... - a_{L-1})$
$\quad + ( - a_1 - a_2 - ... - a_L)$
$\quad + ( - a_2 - ... - a_L)$
$\quad + ( - a_3 - ... - a_L)$
$\quad ...$
$\quad + ( - a_L)$
$)$
$= KL(1 - a_1 - a_2 - ... - a_L) = KL(d/K) = dL$.

Since $(A(1)P(1)-B(1)) = R(1)$, there is no new condition extracted from CASE-4.

CASE-5 (x=-1): We have $(A(-1)P(-1) - B(-1)) = (1 + a_1 - a_2 + a_3 - a_4 + ... -(-1)^L a_L) (q_0 - q_1 + q_2 - q_3 ...+ (-1)^N q_N ) - d((-1)^L + (-1)^{L+1} + ... + (-1)^{N-1} + (-1)^N )$. We also have $R(-1) =$
$K(1 - (1 - a_1) + (1 - a_1 - a_2) - (1 - a_1 - a_2 - a_3) + ... + (-1)^{L-1} (1 - a_1 - a_2 - ... - a_{L-1}) )$
$+ (-1)^N K (( a_1 + a_2 + ... + a_L)$
$- ( a_2 + a_3 ... + a_L)$
$+( a_3 + a_4 + ... + a_L)$
$- ( a_4 + a_5 ... + a_L)$
...
$+ (-1)^L ( a_L) )$

We now consider 4 sub-cases, for $N$ and $L$ each being odd or even.

**Subcase-5.1** *(L is even, N is even)*: $(A(-1)P(-1) - B(-1)) = K(1 + a_1 - a_2 + a_3 ... - a_L) - d$, while $R(-1) = K(a_1 + a_3 + a_5 +...+ a_{L-1}$ $+a_1 + a_3 + a_5 +...+ a_{L-1}) = 2K(a_1 + a_3 + a_5 +...+ a_{L-1})$. So $(A(-1)P(-1) - B(-1)) = K(1 - (d/K) + a_1 - a_2 + a_3 ... - a_L)$. Using the condition from CASE-2, we have $(A(-1)P(-1) - B(-1)) = K(a_1 + a_2 + a_3 ... + a_L + a_1 - a_2 + a_3 ... - a_L) = 2K(a_1 + a_3 + a_5 +...+ a_{L-1})$. $(A(-1)P(-1) - B(-1)) = R(-1)$, so no new condition can be extracted from this Subcase.

**Subcase-5.2** *(L is even, N is odd)*: $P(-1) = B(-1) = 0$, so $(A(-1)P(-1) - B(-1)) = 0$. $R(-1) = K(a_1 + a_3 + a_5 +...+ a_{L-1}$ $- a_1 - a_3 - a_5 -...- a_{L-1}) = 0$. $(A(-1)P(-1) - B(-1)) = R(-1)$, so no new condition can be extracted from this Subcase.

**Subcase-5.3** (L is odd, N is even): $(A(-1)P(-1) - B(-1)) = K(1 + a_1 - a_2 + a_3 - a_4 + ... + a_L)$, while $R(-1) = K(1 - a_2 - a_4 - ... - a_{L-1} + ( a_1 + a_3 + a_5 + ... + a_L )) = K(1 + a_1 - a_2 + a_3 - a_4 + ... + a_L)$. $(A(-1)P(-1) - B(-1)) = R(-1)$, so no new condition can be extracted from this Subcase.

**Subcase-5.4** (L is odd, N is odd): $P(-1) = 0$, so $(A(-1)P(-1) - B(x)) = d$. $R(-1) = K( 1 - a_2 - a_4 - ...- a_{L-1} - ( a_1 + a_3 + ... + a_L ) )$, which as per condition from CASE-4, is $d$. $(A(-1)P(-1) - B(-1)) = R(-1)$, so no new condition is extracted from this Subcase.

Hence, no new condition can be extracted from CASE-5.

Therefore, the condition extracted from all five CASES is $(( q_i = c_i = K$ for all integers $i$ in $[0, L-1] )$ AND $(( a_1 + a_2 + ... + a_{L-1} + a_L ) = (1-(d/K))) )$.

**Hence Proved**

## 4. Polynomial Recurrence Sequences

The first Proof-Method in Theorems 1-4, is easier in the case of linear recurrence sequences (both Homogeneous and non-Homogeneous). This is because when we fix the values of any $L$ successive terms of a linear recurrence sequence, there is only one possible value for the previous term. For example, there is only one possible value for $p_{i-L}$ for any given vector $< p_i , p_{i-1} , ... , p_{i-L+1} >$, and only one possible value of $q_{i-L}$ for any given vector $< q_i , q_{i-1} , ... , q_{i-L+1} >$. Thus, when it is known that a linear recurrence sequence converges to a target rational, there is only one trajectory (hence only one starting point) for the sequence. However, this is not the case in Polynomial Recurrence Sequences, which we denote as follows:

$r_i = c_i$ , for all integers $i$ in $[0,L-1]$ , and

$r_i =$ SUMMATION $( (r_{i-1}{}^{j_1} r_{i-2}{}^{j_2}... r_{i-L}{}^{j_L} a_{j_1,j_2,j_3...j_L} )$, over all $(H+1)^L$ cases of each integer in vector $<j_1, j_2, j_3... j_L>$ being equal to one of the integers in $[0,H]$ ), where $< c_0 , c_1 , c_2 , ... , c_{L-1} >$ and $< a_{1,1,1...1} , ... , a_{H,H,H...H} >$ are given vectors of rationals, where $H$ is some positive integer, and where $r_i$ is the $i^{th}$ term in our sequence.

It is clear that given a starting point $< c_0 , c_1 , c_2 , ... , c_{L-1} >$ and given coefficients $< a_{1,1,1...1} , ... , a_{H,H,H...H} >$ for such a Polynomial Recurrence Sequence, there is only one possible value for $r_L$ , hence only one possible value for $r_{L+1}$ , hence only one possible value for $r_{L+2}$ , and so on. Hence, only one trajectory can be generated in the forward direction.

Instead of being given a starting point, if we are only given that $r_N = r_{N-1} = r_{N-2} = ... = r_{N-L+1} = K$, where $N$ is very large, there are many possible candidate solutions for $r_{N-L}$ , due to the higher degree of terms containing $r_{N-L}$ . Thus, the reverse sequence can potentially have infinite trajectories. This is where the approach described in the second Proof-Method in Theorems 1-4, becomes useful. The second Proof Method expresses the Sequence using a univariate Polynomial equation in $x$ whose coefficients correspond to the terms of the Sequence, and then by analysing cases that cover all real values of $x$, finds a finite number of relationships among the starting points of the sequence, the coefficients of the sequence, and the target rational.

Start of example for our Polynomial Recurrence Sequence:

For example, consider our Polynomial Recurrence Sequence $r_i$ to defined as follows:

$r_0 = c_0$ ;
$r_1 = c_1$ ;
$r_2 = c_2$ ;
$r_i = a_1 (r_{i-1} - r_{i-3}) + a_2 r_{i-2}{}^2 + a_3 r_{i-3} r_{i-1} + d$, for all integers $i \geq 3$,

where, $<c_0, c_1, c_2>$ and $<a_1, a_2, a_3>$ are given vectors of rationals, and where $d$ is a given rational that may be zero or non-zero.

It may be easily proved using an argument similar to the proof of Theorem-3, that $r_i$ can never converge to $0$, if $d \neq 0$.

Define the following for our example (where $N$ tends to infinity):

$A_1(x) = (1 - a_1 x + a_1 x^3)$ ; $\quad P_1(x) = (r_0 + r_1 x + r_2 x^2 + \ldots + r_N x^N)$

$A_2(x) = (-a_2 x^2)$ ; $\quad P_2(x) = (r_0^2 + r_1^2 x + r_2^2 x^2 + \ldots + r_N^2 x^N)$

$A_3(x) = (-a_3 x^3)$ ; $\quad P_3(x) = (r_0 r_2 + r_1 r_3 x + r_2 r_4 x^2 + r_3 r_5 x^3 + \ldots + r_{N-3} r_{N-1} x^{N-3} + r_{N-2} r_N x^{N-2})$

$A_4(x) = (-d)$ ; $\quad P_4(x) = (1 + x + x^2 + \ldots + x^N)$

Now define $R(x) = A_1(x)P_1(x) + A_2(x)P_2(x) + A_3(x)P_3(x) + A_4(x)P_4(x)$. So $R(x) =$

$(r_0 - d) + x(r_1 - a_1 r_0 - d) + x^2(r_2 - a_1 r_1 - a_2 r_0^2 - d) + x^3(r_3 - a_1 r_2 + a_1 r_0 - a_2 r_1^2 - a_3 r_0 r_2 - d) + x^4(r_4 - a_1 r_3 + a_1 r_1 - a_2 r_2^2 - a_3 r_1 r_3 - d) + \ldots + x^{N+1}(-a_1 r_N + a_1 r_{N-2} - a_2 r_{N-1}^2 - a_3 r_{N-2} r_N) + x^{N+2}(a_1 r_{N-1} - a_2 r_N^2) + x^{N+3}(a_1 r_N)$

The coefficients $x^i$ in $R(x)$ for $3 \leq i \leq N$, become zero as they obey the recurrence relation, so we have $R(x) =$

$(r_0 - d) + x(r_1 - a_1 r_0 - d) + x^2(r_2 - a_1 r_1 - a_2 r_0^2 - d) +$
$x^{N+1}(-a_1 r_N + a_1 r_{N-2} - a_2 r_{N-1}^2 - a_3 r_{N-2} r_N) + x^{N+2}(a_1 r_{N-1} - a_2 r_N^2) + x^{N+3}(a_1 r_N)$

Assume $M$ is the minimum integer after which $r_i = K$, for all $i \geq M$. So then $R(x) =$

$(r_0 - d) + x(r_1 - a_1 r_0 - d) + x^2(r_2 - a_1 r_1 - a_2 r_0^2 - d) +$
$x^{N+1}(-a_1 K + a_1 K - a_2 K^2 - a_3 K^2) + x^{N+2}(a_1 K - a_2 K^2) + x^{N+3}(a_1 K)$
$= (r_0 - d) + x(r_1 - a_1 r_0 - d) + x^2(r_2 - a_1 r_1 - a_2 r_0^2 - d) + x^{N+1}(-a_2 K^2 - a_3 K^2) + x^{N+2}(a_1 K - a_2 K^2) + x^{N+3}(a_1 K)$

Our aim is to prove that there exist real Polynomials $P_1(x)$, $P_2(x)$, $P_3(x)$ and $P_4(x)$ such that for every real $x$, $R(x) = A_1(x)P_1(x) + A_2(x)P_2(x) + A_3(x)P_3(x) + A_4(x)P_4(x)$. For this, we consider five cases ($x = 0$, $0 < abs(x) < 1$, $abs(x) > 1$, $x = 1$, and $x = -1$). The union of the conditions obtained in all five cases, will be necessary and sufficient, for $r_i$ to converge to $K$. We proceed to analyse each case in detail.

<u>CASE-1</u> ($x = 0$): $A_1(0)=1$, $P_1(0)=c_0$, $A_2(0)=0$, $P_2(0)=c_0^2$, $A_3(0)=0$, $P_3(0)=c_0 c_2$, $A_4(0)=-d$, $P_4(0)=1$. So $(A_1(x)P_1(x) + A_2(x)P_2(x) + A_3(x)P_3(x) + A_4(x)P_4(x)) = (c_0 - d)$. Even $R(0) = (c_0 - d)$. Hence, no new condition is extracted from CASE-1.

<u>CASE-2</u> ($0 < abs(x) < 1$): $A_1(x)P_1(x) = (1 - a_1 x + a_1 x^3)(r_0 + r_1 x + r_2 x^2 + \ldots + r_{M-1} x^{M-1} + Kx^M/(1-x))$.

$A_2(x)P_2(x) = (-a_2 x^2)(r_0^2 + r_1^2 x + r_2^2 x^2 + \ldots + r_{M-1}^2 x^{M-1} + K^2 x^M/(1-x))$.

$A_3(x)P_3(x) = (-a_3 x^3)(r_0 r_2 + r_1 r_3 x + r_2 r_4 x^2 + r_3 r_5 x^3 + \ldots + r_{M-1} r_{M+1} x^{M-1} + K^2 x^M/(1-x))$.

$A_4(x)P_4(x) = -d(1 + x + x^2 + \ldots + x^{M-1} + x^M/(1-x))$.

$R(x) = (r_0 - d) + x(r_1 - a_1 r_0 - d) + x^2(r_2 - a_1 r_1 - a_2 r_0^2 - d)$, as $N$ tends to infinity.

Since $A_1(x)P_1(x) + A_2(x)P_2(x) + A_3(x)P_3(x) + A_4(x)P_4(x) = R(x)$, we have:

$(1 - a_1 x + a_1 x^3)(r_0 + r_1 x + r_2 x^2 + \ldots + r_{M-1} x^{M-1}) + Kx^M(1 - a_1 x + a_1 x^3)/(1-x) +$
$(-a_2 x^2)(r_0^2 + r_1^2 x + r_2^2 x^2 + \ldots + r_{M-1}^2 x^{M-1}) + K^2 x^M(-a_2 x^2)/(1-x) +$
$(-a_3 x^3)(r_0 r_2 + r_1 r_3 x + r_2 r_4 x^2 + r_3 r_5 x^3 + \ldots + r_{M-1} r_{M+1} x^{M-1}) + K^2 x^M(-a_3 x^3)/(1-x) +$
$(-d)/(1-x)$
$= (r_0 - d) + x(r_1 - a_1 r_0 - d) + x^2(r_2 - a_1 r_1 - a_2 r_0^2 - d)$

We can use the concept of limits [4] to evaluate the value of the expression $\phi$ that will be used in CASE-5. We define $\phi =$

$(r_0 - r_1 + r_2 - r_3 + \ldots + (-1)^{M-1} r_{M-1}) +$
$(-a_2)(r_0^2 - r_1^2 + r_2^2 - r_3^2 + \ldots + (-1)^{M-1} r_{M-1}^2) +$
$(a_3)(r_0 r_2 - r_1 r_3 + r_2 r_4 - r_3 r_5 + \ldots + (-1)^{M-1} r_{M-1} r_{M+1}) +$
$(-d)(1 - 1 + 1 - 1 + \ldots + (-1)^{M-1})$.

From our previous equation $A_1(x)P_1(x) + A_2(x)P_2(x) + A_3(x)P_3(x) + A_4(x)P_4(x) = R(x)$, if we let $x = (-1+h)$, where $h$ is positive and tends to $0$, then we have $\phi =$

$(\quad (r_0 - d) - (1-h)(r_1 - a_1 r_0 - d) + (1-2h)(r_2 - a_1 r_1 - a_2 r_0^2 - d) -$
$(-1)^M(1-Mh)(1/2)(K(1 - a_1(1-h) + a_1(1-3h)) + K^2(-a_2(1-2h)) + K^2(-a_3(1-3h))) \quad )$
$= (\quad (r_0 - d) - (1-h)(r_1 - a_1 r_0 - d) + (1-2h)(r_2 - a_1 r_1 - a_2 r_0^2 - d) -$
$(-1)^M(1/2)(K(1 - 2ha_1) + K^2(-a_2(1-2h)) + K^2(-a_3(1-3h))) \quad )$.

Using the result from CASE-4 that ($K - K^2 a_2 - K^2 a_3 - d = 0$), this means $\phi =$

$($

$(r_0 - d) - (1-h)(r_1 - a_1 r_0 - d) + (1-2h)(r_2 - a_1 r_1 - a_2 r_0^2 - d) -$
$(-1)^M (1/2)(d + K(-2ha_1) + K^2 (2ha_2) + K^2 (3ha_3))$

), where $h$ tends to $0$, and is positive.

Coming back to deriving the conditions for CASE-2, we have:

$(1-x)A_1(x)P_1(x) = (1 - a_1 x + a_1 x^3)(r_0 + (r_1 - r_0)x + (r_2 - r_1)x^2 + ... + (r_{M-1} - r_{M-2})x^{M-1} + (K - r_{M-1})x^M)$.
$(1-x)A_2(x)P_2(x) = (-a_2 x^2)(r_0^2 + (r_1^2 - r_0^2)x + (r_2^2 - r_1^2)x^2 + ... + (r_{M-1}^2 - r_{M-2}^2)x^{M-1} + (K^2 - r_{M-1}^2)x^M)$.
$(1-x)A_3(x)P_3(x) = (-a_3 x^3)(r_0 r_2 + (r_1 r_3 - r_0 r_2)x + (r_2 r_4 - r_1 r_3)x^2 + ... + (r_{M-1} r_{M+1} - r_{M-2} r_M)x^{M-1} + (K^2 - r_{M-1} K)x^M)$.
$(1-x)A_4(x)P_4(x) = -d$
$(1-x)R(x) = (r_0 - d) + (r_1 - a_1 r_0 - r_0)x + (r_2 - a_1 r_1 - a_2 r_0^2 - r_1 + a_1 r_0)x^2 - (r_2 - a_1 r_1 - a_2 r_0^2 - d)x^3$

In $(1-x)R(x) = (1-x)A_1(x)P_1(x) + (1-x)A_2(x)P_2(x) + (1-x)A_3(x)P_3(x) + (1-x)A_4(x)P_4(x)$, we are now able to compare the coefficients of $x^i$ for all integers $i$ in $[0, (M+3)]$. We obtain the following system of equations:

$$
\begin{array}{llllll}
v_3^T f(K, r_{M-1}) & & & & & = 0 \\
v_3^T f(r_{M-1}, r_{M-2}) & + & v_2^T f(K, r_{M-1}) & & & = 0 \\
v_3^T f(r_{M-2}, r_{M-3}) & + & v_2^T f(r_{M-1}, r_{M-2}) & + & v_1^T f(K, r_{M-1}) & = 0 \\
v_3^T f(r_{M-3}, r_{M-4}) & + & v_2^T f(r_{M-2}, r_{M-3}) & + & v_1^T f(r_{M-1}, r_{M-2}) & + & v_0^T f(K, r_{M-1}) & = 0 \\
v_3^T f(r_{M-4}, r_{M-5}) & + & v_2^T f(r_{M-3}, r_{M-4}) & + & v_1^T f(r_{M-2}, r_{M-3}) & + & v_0^T f(r_{M-1}, r_{M-2}) & = 0 \\
\multicolumn{6}{l}{...} \\
v_3^T f(r_3, r_2) & + & v_2^T f(r_4, r_3) & + & v_1^T f(r_5, r_4) & + & v_0^T f(r_6, r_5) & = 0 \\
v_3^T f(r_2, r_1) & + & v_2^T f(r_3, r_2) & + & v_1^T f(r_4, r_3) & + & v_0^T f(r_5, r_4) & = 0 \\
v_3^T f(r_1, r_0) & + & v_2^T f(r_2, r_1) & + & v_1^T f(r_3, r_2) & + & v_0^T f(r_4, r_3) & = 0 \\
v_3^T f(r_0, 0) & + & v_2^T f(r_1, r_0) & + & v_1^T f(r_2, r_1) & + & v_0^T f(r_3, r_2) & = e_3 \\
& & v_2^T f(r_0, 0) & + & v_1^T f(r_1, r_0) & + & v_0^T f(r_2, r_1) & = e_2 \\
& & & & v_1^T f(r_0, 0) & + & v_0^T f(r_1, r_0) & = e_1 \\
& & & & & & v_0^T f(r_0, 0) & = e_0 \\
\end{array}
$$

where $f(r_i, r_{i-1})$ denotes the 3-dimensional vector of variables $= <(r_i - r_{i-1}), (r_i^2 - r_{i-1}^2), (r_i r_{i+2} - r_{i-1} r_{i+1})>^T$, where each of $\{v_0, v_1, v_2, v_3\}$ is a given 3-dimensional vector of constants, with $v_0 = <1, 0, 0>^T$, $v_1 = <-a_1, 0, 0>^T$, $v_2 = <0, -a_2, 0>^T$, $v_3 = <a_1, 0, -a_3>^T$, and, where each of $\{e_0, e_1, e_2, e_3\}$ is a given real constant (i.e. scalar constant), with $e_0 = r_0$, $e_1 = (r_1 - a_1 r_0 - r_0)$, $e_2 = (r_2 - a_1 r_1 - a_2 r_0^2 - r_1 + a_1 r_0)$, and, $e_3 = -(r_2 - a_1 r_1 - a_2 r_0^2 - d)$.

Adding all $(M+4)$ equations, we eliminate variable vector $<r_0, r_1, r_2, ..., r_{M-1}>$ and we get: $(Ka_1 - K^2 a_3 - K^2 a_2 - Ka_1 + K = d)$, which means $(K - a_2 K^2 - a_3 K^2 - d = 0)$, which is the condition extracted from CASE-2.

CASE-3 ($abs(x) \geq 1$): In the equation $R(x) = A_1(x)P_1(x) + A_2(x)P_2(x) + A_3(x)P_3(x) + A_4(x)P_4(x)$, divide both sides with $x^{N+3}$ and let $y=1/x$, so we get:
$y^2 (-a_1 K + a_1 K - a_2 K^2 - a_3 K^2) + y(a_1 K - a_2 K^2) + (a_1 K) =$
$(y^3 - a_1 y^2 + a_1)K/(1-y) + (-a_2 y)K^2/(1-y) + (-a_3 y^2)K^2/(1-y) + (-dy^3)/(1-y)$.
So we have $(1-y)(y^2 (-a_2 K^2 - a_3 K^2) + y(a_1 K - a_2 K^2) + (a_1 K)) = (y^3 - a_1 y^2 + a_1)K + (-a_2 y)K^2 + (-a_3 y^2)K^2 + (-dy^3)$.
So we have $(y^3 (a_2 K^2 + a_3 K^2) + y^2 (-a_3 K^2 - a_1 K) + y(-a_2 K^2) + (a_1 K)) =$
$(y^3 - a_1 y^2 + a_1)K + (-a_2 y)K^2 + (-a_3 y^2)K^2 + (-dy^3) = (y^3 (K - d) + y^2 (-a_3 K^2 - a_1 K) + y(-a_2 K^2) + (a_1 K))$. Comparing coefficients, we get $(K - a_2 K^2 - a_3 K^2 - d = 0)$, which is the condition extracted from CASE-3.

CASE-4 ($x=1$): $R(1) = (r_0 - d) + (r_1 - a_1 r_0 - d) + (r_2 - a_1 r_1 - a_2 r_0^2 - d) + (-a_2 K^2 - a_3 K^2) + (a_1 K - a_2 K^2) + (a_1 K) =$
$(r_0 - 3d) + (r_1 - a_1 r_0) + (r_2 - a_1 r_1 - a_2 r_0^2) + 2a_1 K - 2a_2 K^2 - a_3 K^2$

We also have:
$A_1(1)P_1(1) = (1 - a_1 + a_1)(r_0 + r_1 + r_2 + ... + r_N) = (r_0 + r_1 + r_2 + ... + r_N)$
$A_2(1)P_2(1) = (-a_2)(r_0^2 + r_1^2 + r_2^2 + ... + r_N^2)$
$A_3(1)P_3(1) = (-a_3)(r_0 r_2 + r_1 r_3 + r_2 r_4 + r_3 r_5 + ... + r_{N-3} r_{N-1} + r_{N-2} r_N)$
$A_4(1)P_4(1) = (-d)(N)$

As the value of $N$ tends to infinity, the value of $A_1(1)P_1(1) + A_2(1)P_2(1) + A_3(1)P_3(1) + A_4(1)P_4(1) =$
$N(K - K^2 a_2 - K^2 a_3 - d)$. Since the value of $R(1)$ is finite, we need $(K - K^2 a_2 - K^2 a_3 - d) = 0$. We also need $R(1) = 0$, so we need $((c_0 - 3d) + (c_1 - a_1 c_0) + (c_2 - a_1 c_1 - a_2 c_0^2) + 2a_1 K - 2a_2 K^2 - a_3 K^2) = 0$. Hence, the condition extracted from CASE-4 is $((K - K^2 a_2 - K^2 a_3 - d = 0)$ AND $(c_0 - 3d + c_1 - a_1 c_0 + c_2 - a_1 c_1 - a_2 c_0^2 + 2a_1 K - 2a_2 K^2 - a_3 K^2 = 0))$.

<u>CASE-5 $(x=-1)$</u>: $R(-1) = (c_0 - d) + (-1)(c_1 - a_1 c_0 - d) + (c_2 - a_1 c_1 - a_2 c_0^2 - d) + (-1)^{N+1}(-a_2 K^2 - a_3 K^2) + (-1)^{N+2}(a_1 K - a_2 K^2) + (-1)^{N+3}(a_1 K) = c_0 - d - c_1 + a_1 c_0 + c_2 - a_1 c_1 - a_2 c_0^2 + (-1)^{N+1}((-a_2 K^2 - a_3 K^2) - (a_1 K - a_2 K^2) + (a_1 K)) =$
$c_0 - c_1 + c_2 - d + a_1 c_0 - a_1 c_1 - a_2 c_0^2 + (-1)^N (a_3 K^2)$. We also have:

$A_1(-1)P_1(-1) = (r_0 - r_1 + r_2 - r_3 + \ldots + (-1)^N r_N)$
$A_2(-1)P_2(-1) = (-a_2)(r_0^2 - r_1^2 + r_2^2 - r_3^2 + \ldots + (-1)^N r_N^2)$
$A_3(-1)P_3(-1) = (a_3)(r_0 r_2 - r_1 r_3 + r_2 r_4 - r_3 r_5 + \ldots + (-1)^{N-2} r_{N-2} r_N)$
$A_4(-1)P_4(-1) = (-d)(1-1+1-1+\ldots+(-1)^N)$

We have earlier defined $\phi = (r_0 - r_1 + r_2 - r_3 + \ldots + (-1)^{M-1} r_{M-1}) +$
$(-a_2)(r_0^2 - r_1^2 + r_2^2 - r_3^2 + \ldots + (-1)^{M-1} r_{M-1}^2) +$
$(a_3)(r_0 r_2 - r_1 r_3 + r_2 r_4 - r_3 r_5 + \ldots + (-1)^{M-1} r_{M-1} r_{M+1}) +$
$(-d)(1-1+1-1+\ldots+(-1)^{M-1})$. We have given also an approach how to calculate $\phi$, using the concept of limits, in CASE-2.

We now have 4 Subcases:

Subcase-4.1 (N is even AND M is even): $R(-1) = c_0 - c_1 + c_2 - d + a_1 c_0 - a_1 c_1 - a_2 c_0^2 + a_3 K^2$. And $A_1(-1)P_1(-1) + A_2(-1)P_2(-1) + A_3(-1)P_3(-1) + A_4(-1)P_4(-1) = \phi + (K - K^2 a_2 + K^2 a_3 - d) = \phi + 2K^2 a_3$.

Subcase-4.2 (N is even AND M is odd): $R(-1) = c_0 - c_1 + c_2 - d + a_1 c_0 - a_1 c_1 - a_2 c_0^2 + a_3 K^2$. And $A_1(-1)P_1(-1) + A_2(-1)P_2(-1) + A_3(-1)P_3(-1) + A_4(-1)P_4(-1) = \phi + 0 = \phi$.

Subcase-4.3 (N is odd AND M is even): $R(-1) = c_0 - c_1 + c_2 - d + a_1 c_0 - a_1 c_1 - a_2 c_0^2 - a_3 K^2$. And $A_1(-1)P_1(-1) + A_2(-1)P_2(-1) + A_3(-1)P_3(-1) + A_4(-1)P_4(-1) = \phi + 0 = \phi$.

Subcase-4.4 (N is odd AND M is odd): $R(-1) = c_0 - c_1 + c_2 - d + a_1 c_0 - a_1 c_1 - a_2 c_0^2 - a_3 K^2$. And $A_1(-1)P_1(-1) + A_2(-1)P_2(-1) + A_3(-1)P_3(-1) + A_4(-1)P_4(-1) = \phi - (K - K^2 a_2 + K^2 a_3 - d) = \phi - 2K^2 a_3$.

So the condition extracted from CASE-5 is:
$((c_0 - c_1 + c_2 - d + a_1 c_0 - a_1 c_1 - a_2 c_0^2 + a_3 K^2 = \phi + 2K^2 a_3)$ OR
$(c_0 - c_1 + c_2 - d + a_1 c_0 - a_1 c_1 - a_2 c_0^2 + a_3 K^2 = \phi)$ OR
$(c_0 - c_1 + c_2 - d + a_1 c_0 - a_1 c_1 - a_2 c_0^2 - a_3 K^2 = \phi)$ OR
$(c_0 - c_1 + c_2 - d + a_1 c_0 - a_1 c_1 - a_2 c_0^2 - a_3 K^2 = \phi - 2K^2 a_3))$,

which is equivalent to the condition:

$(\quad((c_0 - c_1 + c_2 - d + a_1 c_0 - a_1 c_1 - a_2 c_0^2 = \phi + K^2 a_3)$ if M is even$)$
OR
$((c_0 - c_1 + c_2 - d + a_1 c_0 - a_1 c_1 - a_2 c_0^2 = \phi - K^2 a_3)$ if M is odd$)\quad)$.

Now, taking the value of $\phi$ from CASE-2, we have

If M is even: $\phi = ((r_0 - d) - (1-h)(r_1 - a_1 r_0 - d) + (1-2h)(r_2 - a_1 r_1 - a_2 r_0^2 - d) -$
$(1/2)(d + K(-2ha_1) + K^2(2ha_2) + K^2(3ha_3)) \quad)$

If M is odd: $\phi = ((r_0 - d) - (1-h)(r_1 - a_1 r_0 - d) + (1-2h)(r_2 - a_1 r_1 - a_2 r_0^2 - d) +$
$(1/2)(d + K(-2ha_1) + K^2(2ha_2) + K^2(3ha_3)) \quad)$

So our condition extracted from CASE-5 is:
If M is even:
$(c_0 - c_1 + c_2 - d + a_1 c_0 - a_1 c_1 - a_2 c_0^2 - K^2 a_3) =$
$(\quad (c_0 - d) - (1-h)(c_1 - a_1 c_0 - d) + (1-2h)(c_2 - a_1 c_1 - a_2 c_0^2 - d) -$
$(1/2)(d + K(-2ha_1) + K^2(2ha_2) + K^2(3ha_3)) \quad)$

If M is odd:
$(c_0 - c_1 + c_2 - d + a_1 c_0 - a_1 c_1 - a_2 c_0^2 + K^2 a_3) =$
$(\quad (c_0 - d) - (1-h)(c_1 - a_1 c_0 - d) + (1-2h)(c_2 - a_1 c_1 - a_2 c_0^2 - d) +$
$(1/2)(d + K(-2ha_1) + K^2(2ha_2) + K^2(3ha_3)) \quad)$

So our condition extracted from CASE-5 is:
If M is even: $(-K^2 a_3) = (\quad h(c_1 - a_1 c_0 - d) - 2h(c_2 - a_1 c_1 - a_2 c_0^2 - d) -$
$(1/2)(d + K(-2ha_1) + K^2(2ha_2) + K^2(3ha_3)) \quad)$

If M is odd: $(+K^2 a_3) = (\quad h(c_1 - a_1 c_0 - d) - 2h(c_2 - a_1 c_1 - a_2 c_0^2 - d) +$

$$(1/2)( d + K( -2ha_1 ) + K^2 (2ha_2) + K^2 ( 3ha_3) ) \quad )$$

So our condition extracted from CASE-5 is:

If $M$ is even: $( -K^2 a_3 - d/2 ) = h( \quad (c_1 - a_1 c_0 + d) - 2(c_2 - a_1 c_1 - a_2 c_0^2) -$
$\quad\quad\quad\quad\quad\quad\quad\quad ( K(-a_1) + K^2(a_2) + K^2(3a_3/2) ) \quad )$

If $M$ is odd: $( +K^2 a_3 - d/2 ) = h( \quad (c_1 - a_1 c_0 + d) - 2(c_2 - a_1 c_1 - a_2 c_0^2) +$
$\quad\quad\quad\quad\quad\quad\quad\quad ( K(-a_1) + K^2(a_2) + K^2(3a_3/2) ) \quad )$

Since $h$ tends to zero, we get our condition extracted from CASE-5 as:

$( \quad\quad (K^2 a_3 + d/2 = 0) \text{ OR } (K^2 a_3 - d/2 = 0) \quad )$

Hence, the necessary and sufficient boolean condition $V$ extracted from all five CASES, for our example, is:

(

$(K - a_2 K^2 - a_3 K^2 - d = 0)$ **AND**

$(c_0 - 3d + c_1 - a_1 c_0 + c_2 - a_1 c_1 - a_2 c_0^2 + 2a_1 K - 2a_2 K^2 - a_3 K^2 = 0 )$ **AND**

$(( 2K^2 a_3 + d = 0) \textbf{ OR } ( 2K^2 a_3 - d = 0))$

).

( $V$ is TRUE ) ↔ ( The Polynomial Recurrence $r_i$ in our example converges to $K$ ).

<u>End of example for our Polynomial Recurrence Sequence</u>

## 5. Conclusion

We showed that it is easy to determine whether or not a given linear recurrence sequence can converge to a given target rational. We also showed that it is not straightforward to determine the same for polynomial recurrence sequences, because given that the sequence converges, there are potentially infinite reverse sequences that can be generated, due to the higher degree of terms in the definition of the Polynomial Recurrence Sequence.

We then presented an efficient polynomial-time approach to determine whether or not a given Polynomial Recurrence Sequence can converge to a given target rational. The basic idea of the approach is to construct a univariate polynomial equation in $x$, whose coefficients correspond to the terms of the Sequence. The approach then obtains a condition by analyzing 5 CASES ($x = 0$, $0 < abs(x) < 1$, $abs(x) > 1$, $x = 1$, and, $x = -1$). The obtained condition is necessary and sufficient for the convergence of the given Polynomial Recurrence Sequence, because it covers all possible real values of $x$.

## 6. Future Work

There are two interesting areas of future work. First, can this approach be used to obtain the generic closed form expression for the terms of Polynomial Recurrence Sequence? Second, can this approach be extended to be used for Recurrence Sequences, where the relation is not Polynomial, but sinusoidal or exponential or rational-Polynomial?

**About the Author**

I, Deepak Ponvel Chermakani, wrote this paper out of my own interest and initiative, during my spare time. In Sep-2010, I completed a fulltime one year Master Degree in *Operations Research with Computational Optimization*, from University of Edinburgh UK (*www.ed.ac.uk*). In Jul-2003, I completed a fulltime four year Bachelor Degree in *Electrical and Electronic Engineering*, from Nanyang Technological University Singapore (*www.ntu.edu.sg*). In Jul-1999, I completed fulltime high-schooling from National Public School in Bangalore in India.